\documentclass{jpp}
\usepackage{subeqn}
\usepackage{epsfig}

\let\realverbatim\verbatim
\let\realendverbatim\endverbatim
\renewenvironment{verbatim}
  {\par\addvspace{6pt plus 1pt minus 2pt}\realverbatim}
  {\realendverbatim\addvspace{6pt plus 1pt minus 2pt}}

\ifprodtf \else
  \checkfont{eurm10}
  \iffontfound
    \IfFileExists{upmath.sty}
      {\typeout{^^JFound AMS Euler Roman fonts on the system,
                   using the 'upmath' package.^^J}%
       \usepackage{upmath}}
      {\typeout{^^JFound AMS Euler Roman fonts on the system, but you
                   don't seem to have the}%
       \typeout{'upmath' package installed. JPP.cls can take advantage
                 of these fonts,^^Jif you use 'upmath' package.^^J}%
       \providecommand\umu{\umu}%
      }
  \else
    \providecommand\umu{\mu}%
  \fi
\fi

\ifprodtf \else
  \checkfont{msam10}
  \iffontfound
    \IfFileExists{amssymb.sty}
      {\typeout{^^JFound AMS Symbol fonts on the system, using the
                'amssymb' package.^^J}%
       \usepackage{amssymb}%

      }{}
  \else
  \fi
\fi

\ifprodtf \else
  \IfFileExists{amsbsy.sty}
    {\typeout{^^JFound the 'amsbsy' package on the system, using it.^^J}%
     \usepackage{amsbsy}}
    {}
\fi

\newcommand{\eqsto}[2]{Eqs. (\ref{#1}) to (\ref{#2})}

\newcommand{\be}{\begin{equation}}
\newcommand{\ee}{\end{equation}}

\newcommand{\eqa}{\begin{eqnarray}}
\newcommand{\eqe}{\end{eqnarray*}}
\newcommand{\eqnu}{\begin{eqnarray}}
\newcommand{\eqne}{\end{eqnarray}}

\newcommand{\eqs}[2]{Eqs. (\ref{#1}) \& (\ref{#2})}

\newcommand{\eq}[1]{Eq. (\ref{#1})}

\newcommand{\eeq}{\end{eqnarray}}

\newdefinition{definition}[theorem]{Definition}

\title[Journal of Plasma Physics]
{Linear modes in the  partially ionized  heliosphere plasma}

\author[D. Shaikh]
{M. \ls E. \ls K\ls E\ls L\ls L\ls U\ls M\ls$^{1}$ 
\thanks{\tt mek@calhoun.edu}\ls and
D\ls A\ls S\ls T\ls G\ls E\ls E\ls R \ns S\ls H\ls A\ls I\ls K\ls H\ls$^{2}$
\thanks{\tt Email:dastgeer.shaikh@uah.edu} }
\affiliation{$^1$ Division of Mathematics,
Calhoun Community College, \\P. O. Box 2216
Decatur, AL 35609-2216\\
$^2$Department of Physics and \\
Center for Space Physics and Aeronomic Research (CSPAR),\\
University of Alabama at Huntsville, Huntsville, AL 35805. USA.}
\date{Nov 2 2010, Accepted on Dec 4, 2010}
\pubyear{2010} 
\volume{00}
\part{0}
\pagerange{\pageref{firstpage}--\pageref{lastpage}}
\doi{S0963548301004989}

\begin{document}

\label{firstpage}
\maketitle

\begin{abstract}
The heliosphere is predominantly a partially ionized plasma that
consists of electrons, ions and significant neutral atoms.  Nonlinear
interactions amongst these species take place through direct collision
or charge exchange processes. These interactions modify linear and non
linear properties of the plasma. In this work, we develop a one
dimensional linear theory to investigate linear instabilities in such
system.  In our model, the electrons and ions are described by a
single fluid compressible magnetohydrodynamic (MHD) model and are
coupled self-consistently to the neutral fluid via compressible
hydrodynamic equations. The coupling is mediated by the charge
exchange process.  Based on our self-consistent analysis, we find that
the charge exchange coupling is more effective at larger length
scales, and Alfv\'en waves are not affected by the charge exchange
coupling. By contrast, the fast and slow waves are driven linearly
unstable.
\end{abstract}


\section{Introduction}
Magnetohydrodynamic (MHD) plasma admits fundamental modes in the form
of Alfv\'en, fast and slow waves. These modes are ubiquitously
observed in many laboratory [\cite{lab}], space
[\cite{Gekelman,dastgeer3}] and astrophysical plasmas
[\cite{balsara}].  The Alfv\'en, fast and slow waves are essentially
electromagnetic oscillations in a plasma that propagate along an
ambient or guide magnetic field. Their propagation characteristic
depends on the orientation of their wave vector relative to the
velocity and magnetic field fluctuations. For instance, wave vector
associated with Alfv\'en waves is orthogonal to the velocity field
fluctuations whereas the fast and slow waves have the parallel wave
vector relative to the velocity field fluctuations
[\cite{dastgeer3,dastgeer4}].

The propagation properties of the Alfv\'en, fast and slow waves have
long been known and well studied in the MHD plasma (in isolation). But
their evolution in the presence of complex interactions or other
species, such as neutral atoms/particles, is far less explored because
of untractable nature of analytic as well as numerics. In the context
of partially ionized (consisting of ions, electrons and significant
neutrals) space and astrophysical plasmas, these waves interact with
the neutral gas and govern numerous properties.  Kulsrud \& Pearce
(1969) in the context of cosmic ray propagation noted that the
interaction of a neutral gas and plasma can damp Alfv\'en waves.
Neutrals interacting with plasma via a relative drag process results
in ambipolar diffusion [Oishi \& Mac Low 2006]. Ambipolar diffusion
plays a crucial role in the dynamical evolution of the near solar
atmosphere, interstellar medium, and molecular clouds and star
formation.  Oishi \& Mac Low (2006) investigated that the ambipolar
diffusion sets a characteristic mass scale in molecular clouds. They
found presence of structures below the ambipolar diffusion scale
because of the propagation of compressive slow mode MHD waves at
smaller scales. Leake et al (2005) showed that the lower chromosphere
contains neutral atoms, the existence of which greatly increases the
efficiency of wave damping due to collisional friction momentum
transfer.  They noted that Alfv\'en waves with frequencies above 0.6Hz
are completely damped and frequencies below 0.01 Hz are
unaffected. They undertook a quantitative comparative study of the
efficiency of the role of (ion-neutral) collisional friction, viscous
and thermal conductivity mechanisms in damping MHD waves in different
parts of the solar atmosphere.  It was pointed out by the authors that
a correct description of MHD wave damping requires the consideration
of all energy dissipation mechanisms through the inclusion of the
appropriate terms in the generalized Ohm’s law, the momentum, energy
and induction equations.  Padoan et al (2000) calculated frictional
heating by ion-neutral (or ambipolar) drift in turbulent magnetized
molecular clouds and showed that the ambipolar heating rate per unit
volume depends on field strength for constant rms Mach number of the
flow, and on the Alfv\'enic Mach number.

\begin{figure}[t]
\includegraphics[width=11cm]{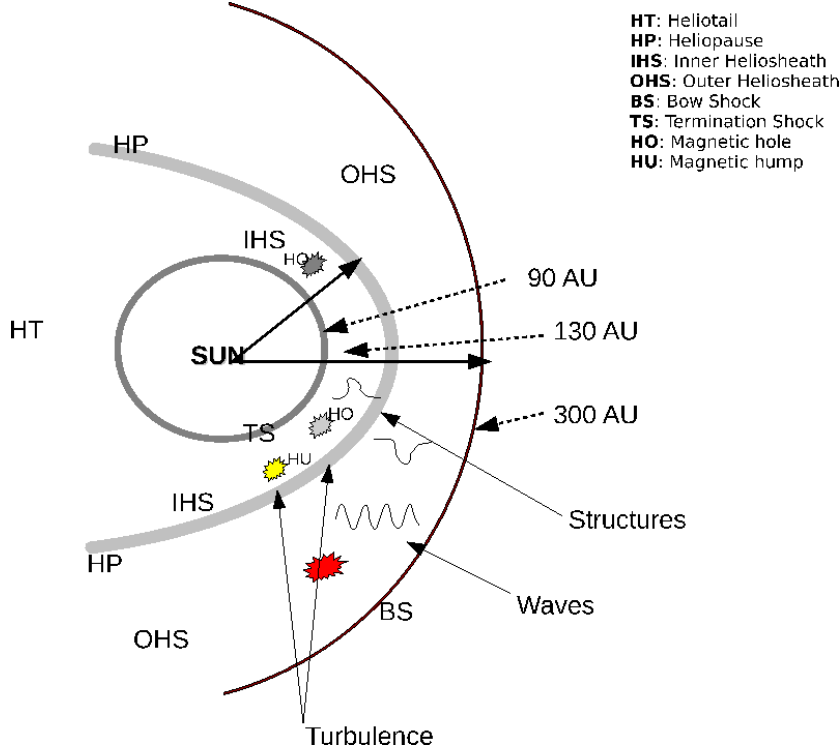}
\caption{Schematic overview of different regions in the global heliosphere. 
The solar wind emanating from the Sun propagates outward and interacts
with partially ionized interstellar gas predominantly via charge
exchange, and creates pick up ions (PUIs). At the termination shock
(TS), the supersonic SW decelerated, heated, compressed becoming
subsonic, in the heliosheath, and again interacts with interstellar
neutrals via charge exchange before it reaches heliopause (HP). The
subsonic SW flows down into the heliotail.  During its journey from
the Sun to the HP, the solar wind plasma develops multitude of length
and time scales that interact with the partially ionized interstellar
gas, TS, and nonlinear structures develop in a complex manner. }
\label{fig0}
\end{figure}

In the context of molecular clouds, the role of ion-neutral collision
has been investigated by Balsara (1996). It was found that momentum of
the plasma is governed predominantly by the slow comoving neutrals
that tend to dissipate Cloud's magnetic field. Balsara (1996)
described a linear analysis by considering neutral momentum in the
presence of gravitational field of the Clouds that dominates the
plasma momenta because of their higher masses. Hence they ignored the
inertial terms in plasma momentum equation. Balsara (1996) reported
that in super Alfv\'enic regime, the slow waves propagate without
significant damping near short wavelengths, while the fast and
Alfv\'en waves undergo rapid damping.

In addition to the neutral gas interacting collisionally with the
plasma, it undergoes charge exchange process in the heliosphere and in
the local interstellar medium (ISM)
[\cite{zank1999,dastgeer}]. Various regions of interactions in the
heliosphere are depicted in Fig (1).  The charge exchange process
knocks off electron from a neutral and makes it a ion. This electron
is captured by the plasma ion which then turns into a neutral atom.
This process conserves the density and but not momentum and energy of
the plasma protons and H neutral gas.  In the local interstellar
medium (LISM), just beyond the heliosphere, the low density plasma and
neutral hydrogen (H) gas are coupled primarily through the process of
charge exchange. Thus on sufficiently large temporal and spatial
scales, a partially ionized plasma is typically regarded as
equilibrated.  Nonetheless it remains to be seen as to what happens to
these MHD modes in the presence of charge exchange process, both in
the inner and outer helioshpere regions. We address this question in
this paper based on a linear theory that describes the coupled
plasma-neutral system mediated by the charge exchange process within
the realm of heliosphere.  While obviously of interest to the solar
wind-LISM interaction [\cite{zank1999,dastgeer}], this work will have
interesting implications for any shock, structures, or turbulent
fluctuations embedded in a partially ionized medium, ranging from
supernova shocks (SNR) to cometary bow shocks.

In section 2, we describe the governing equations of the coupled
plasma-neutral system in the heliosphere plasma. Importance of the
charge exchange is discussed. Section 3 deals with the linear theory
of MHD modes in the presence of charge exchange. Here the plasma and
neutral fluids are coupled via charge exchange.  Because of the
analytic complexities, we focus only on one dimensional calculation.
The solution of linear dispersion relation for the coupled
plasma-neutral system is described in section 4.  In section 5, we
summarize our results.

\section{Model equations}
Within the paradigm of the heliosphere plasma, the plasma and neutral
fluid remain close to thermal equilibirium and behave as Maxwellian
fluids.  They are coupled through charge exchange forces. There may
additionally be direct collision between the two components. We
however ignore it for our current calculation.  In general, the fluid
model describing nonlinear turbulent processes, in the presence of
charge exchange and collision, can be described in terms of the plasma
density ($\rho$), velocity (${\bf U}$), magnetic field (${\bf B}$),
pressure ($P$) components according to the conservative form

\be
\label{mhd}
 \frac{\partial {\bf F}_p}{\partial t} + \nabla \cdot {\bf Q}_p={\cal Q}_{p,n},
\ee
where,
\[{\bf F}_p=
\left[ 
\begin{array}{c}
\rho  \\
\rho {\bf U}  \\
{\bf B} \\
e_p
  \end{array}
\right], 
{\bf Q}_p=
\left[ 
\begin{array}{c}
\rho {\bf U}  \\
\rho {\bf U} {\bf U}+ \frac{P}{\gamma-1}+\frac{B^2}{8\pi}-{\bf B}{\bf B} \\
{\bf U}{\bf B} -{\bf B}{\bf U}\\
e_p{\bf U}
-{\bf B}({\bf U} \cdot {\bf B})
  \end{array}
\right],\]
\[ {\cal Q}_{p,n}=
\left[ 
\begin{array}{c}
0  \\
{\bf Q}_{M,p,n} + {\bf{F}}_{p,n}   \\
0 \\
Q_{E,p,n} + {\bf U} \cdot {\bf{F}}_{p,n}
  \end{array}
\right]
\] 
and
\[ e_p=\frac{1}{2}\rho U^2 + \frac{P}{\gamma-1}+\frac{B^2}{8\pi}.\]
Note the presence of the source terms $Q$ that couple the plasma
self-consistently to the neutral gas.  The above set of plasma
equations is supplimented by $\nabla \cdot {\bf B}=0$ and is coupled
self-consistently to the neutral density ($N$), velocity (${\bf
  V}$) and pressure ($P_n$) through a set of hydrodynamic fluid
equations,
\be
\label{hd}
 \frac{\partial {\bf F}_n}{\partial t} + \nabla \cdot {\bf Q}_n={\cal Q}_{n,p},
\ee
where,
\[{\bf F}_n=
\left[ 
\begin{array}{c}
N  \\
N {\bf V}  \\
e_n
  \end{array}
\right], 
{\bf Q}_n=
\left[ 
\begin{array}{c}
N {\bf V}  \\
N {\bf V} {\bf V}+ \frac{P_n}{\gamma-1} \\
e_n{\bf V}
  \end{array}
\right],\]
\[{\cal Q}_{n,p}=
\left[ 
\begin{array}{c}
0  \\
{\bf Q}_{M,n,p} + {\bf{F}}_{n,p}   \\
Q_{E,n,p} + {\bf V} \cdot {\bf{F}}_{n,p}
  \end{array}
\right],
\] 
\[e_n= \frac{1}{2}NV^2 + \frac{P_n}{\gamma-1}.\]
Equations (\ref{mhd}) to (\ref{hd}) form an entirely self-consistent
description of the coupled  plasma-neutral turbulent fluid in a
partially ionized medium.

Several points are worth noting.  The charge-exchange momentum sources
in the plasma and the neutral fluids, i.e.  \eqs{mhd}{hd}, are
described respectively by terms ${\bf Q}_{M,p,n}({\bf U},{\bf V},\rho,
N, T_n, T_p)$ and ${\bf Q}_{M,n,p}({\bf V},{\bf U},\rho, N, T_n,
T_p)$. These expressions are described in Pauls et al (1995) and
Shaikh \& Zank (2008, 2010).  A swapping of the plasma and the neutral
fluid velocities in this representation corresponds, for instance, to
momentum changes (i.e. gain or loss) in the plasma fluid as a result
of charge exchange with the neutral atoms (i.e. ${\bf Q}_{M,p,n}({\bf
U},{\bf V},\rho, N, T_n, T_p)$ in Eq. (\ref{mhd})). Similarly,
momentum change in the neutral fluid by virtue of charge exchange with
the plasma ions is described by ${\bf Q}_{M,n,p}({\bf V},{\bf U},\rho,
N, T_n, T_p)$ in Eq. (\ref{hd}).

The underlying coupled fluid model can be non-dimensionalized
straightforwardly using a typical scale-length ($\ell_0$), density
($\rho_0$) and velocity ($v_0$). The normalized plasma density,
velocity, energy and the magnetic field are respectively;
$\bar{\rho} = \rho/\rho_0, \bar{\bf U}={\bf U}/v_0,
\bar{P}_p=P_p/\rho_0v_0^2, \bar{\bf B}={\bf B}/v_0\sqrt{\rho_0}$. The
corresponding neutral fluid quantities are $\bar{N} =
N/\rho_0, \bar{\bf V}={\bf V}/v_0,
\bar{P}_n=P_n/\rho_0v_0^2$. The momentum and the energy
charge-exchange terms, in the normalized form, are respectively
$\bar{\bf Q}_m={\bf Q}_m \ell_0/\rho_0v_0^2, \bar{Q}_e=Q_e
\ell_0/\rho_0v_0^3$. The non-dimensional temporal and spatial
length-scales are $\bar{t}=tv_0/\ell_0, \bar{\bf x}={\bf
  x}/\ell_0$. Note that we have removed bars from the set of
normalized coupled model equations (\ref{mhd}) \& (\ref{hd}).

\begin{figure}[t]
\includegraphics[width=11cm]{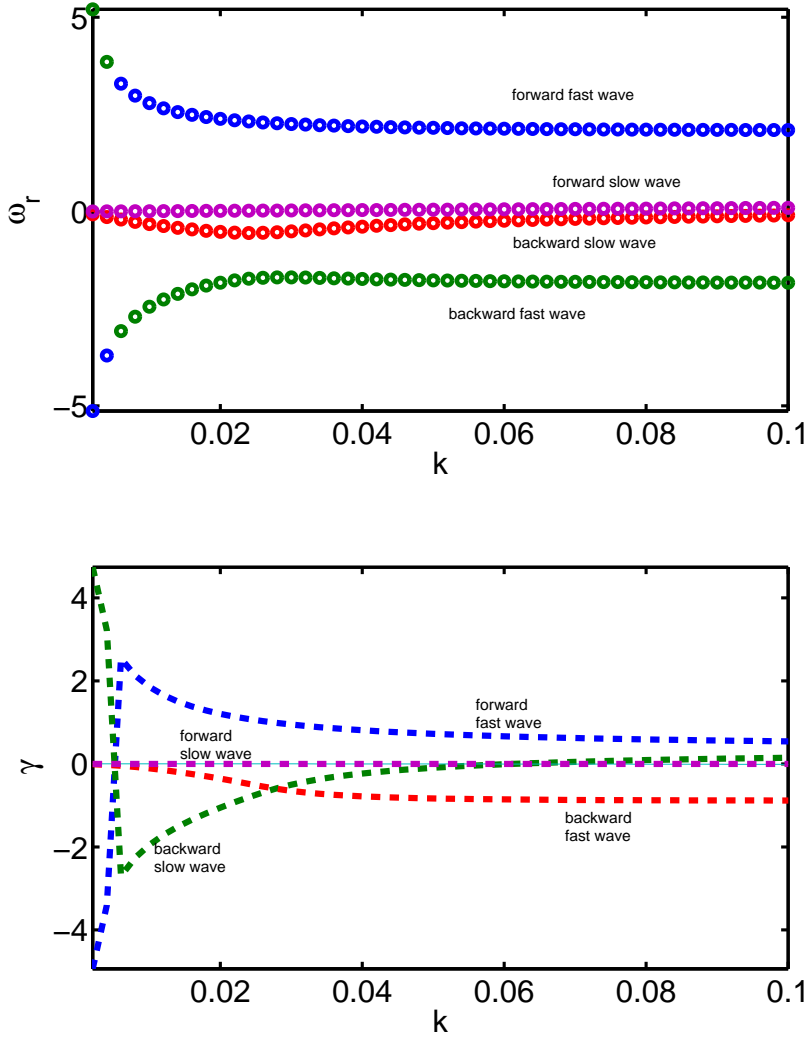}
\caption{Solution of linear dispersion relation in the smaller $k$
(large scale) regime characterizes different MHD and hydrodynamic
modes in the coupled plasma-neutral system. Upper panel describes real
part of the frequency where forward/backward fast and slow waves are
shown.  Growth and damping of linear instability are shown in the
lower panel, where $\gamma=0$ root corresponds to the sound wave. The
latter is stable at lower $k$'s.}
\label{fig1}
\end{figure}

\section{Linear theory}
To avoid mathematical complications associated with the coupled plasma
neutral system, we develop a one dimensional linear theory to
investigate growth/damping of fundamental MHD modes in the heliosphere
plasma.  One dimensional theory offers not only a better insight into
the complex (charge exchange) interaction processes, but it also is
easier (compared to two dimensional treatment) to deal with
analytically.  Furthermore, it describes dominant features of the
characteristic modes of the coupled plasma-neutral system.  In our
linear theory, the charge exchange term $Q_m$ in the momentum equation is
proportional to $\nu \rho (U -V)$ [Florinski et al 2005] where $\nu$
is the charge exchange frequency.  This is a simpler form of the
charge exchange force. For more complicated structure of the charge
exchange force, we employ higher order terms for the nonlinear
simulations. This is described in Shaikh \& Zank (2010) where
nonlinear stage of mode coupling instability in two dimension was
investigated. Our results, to be deduced below, help explain some of
the features of the nonlinear simulations described in Shaikh \& Zank
(2010).
The one-dimensional plasma equations in the normalized form are described as;
\be
\label{plasma1}
\frac{\partial \rho}{\partial t} + 
\frac{\partial }{\partial x} (\rho U)= 0,
\ee
\be
\label{plasma2}
\rho \frac{\partial U}{\partial t} + 
\rho U \frac{\partial U}{\partial x} 
+ \frac{\partial P}{\partial x} = -\nu \rho (U -V),
\ee
\be
\label{plasma3}
\frac{\partial P}{\partial t} + 
 U \frac{\partial P}{\partial x} 
+ \gamma P \frac{\partial U}{\partial x} = 
\nu \rho \left[ \frac{\gamma-1}{2} (U -V)^2 +v_{th}^2 - \frac{P}{2\rho} \right],
\ee
\be
\label{plasma4}
\frac{\partial B}{\partial t} = B \frac{\partial U}{\partial x}
-U \frac{\partial B}{\partial x} - B \frac{\partial U}{\partial x}.
\ee
Here $\nu=\sigma U^* N$ is charge exchange frequency that depends on
the charge exchange cross section ($\sigma$) and the relative
difference in the speed of neutral and plasma components.  The plasma
thermal speed is $v_{th} = \sqrt{kT_i/m_i}$ and $\gamma$ is the
specific heat ratio (it is 5/3 in our model).  Note that the sign of
the momentum charge exchange force is negative.  This means that
charge exchange process tends to decrease plasma momentum relative to
the neutral fluid. The magnetic field is unaffected by the charge
exchange process and it continues to be frozen into the plasma
component of the flow.  Similar equations for the neutral fluid
density ($N$), velocity ($V$) and pressure ($P_n$)  can be described as follows.
\be
\label{neut1}
\frac{\partial N}{\partial t} + 
\frac{\partial }{\partial x} (N V)= 0,
\ee
\be
\label{neut2}
N \frac{\partial V}{\partial t} + 
N V \frac{\partial V}{\partial x} 
+ \frac{\partial P_n}{\partial x} = +\nu \rho (U -V),
\ee
\be
\label{neut3}
 \frac{\partial P_n}{\partial t} + 
 V \frac{\partial P_n}{\partial x} 
+ \gamma P_n \frac{\partial V}{\partial x} = 
\nu \rho \left[ \frac{\gamma-1}{2} (U -V)^2 +v_{th}^2 - \frac{P_n}{2N} \right].
\ee
Unlike plasma, the momentum charge exchange force corresponding to the
neutral fluid is positive. This leads to an enhancement in the neutral
momentum relative to the plasma fluid. The rate of decay or
enhancement of both the plasma and neutral momentum also depends on
the relative speed between the two species. It is evident from
\eqsto{plasma1}{neut3} that the plasma and neutral fluids are
coupled through charge exchange sources.
 
\begin{figure}[t]
\includegraphics[width=11cm]{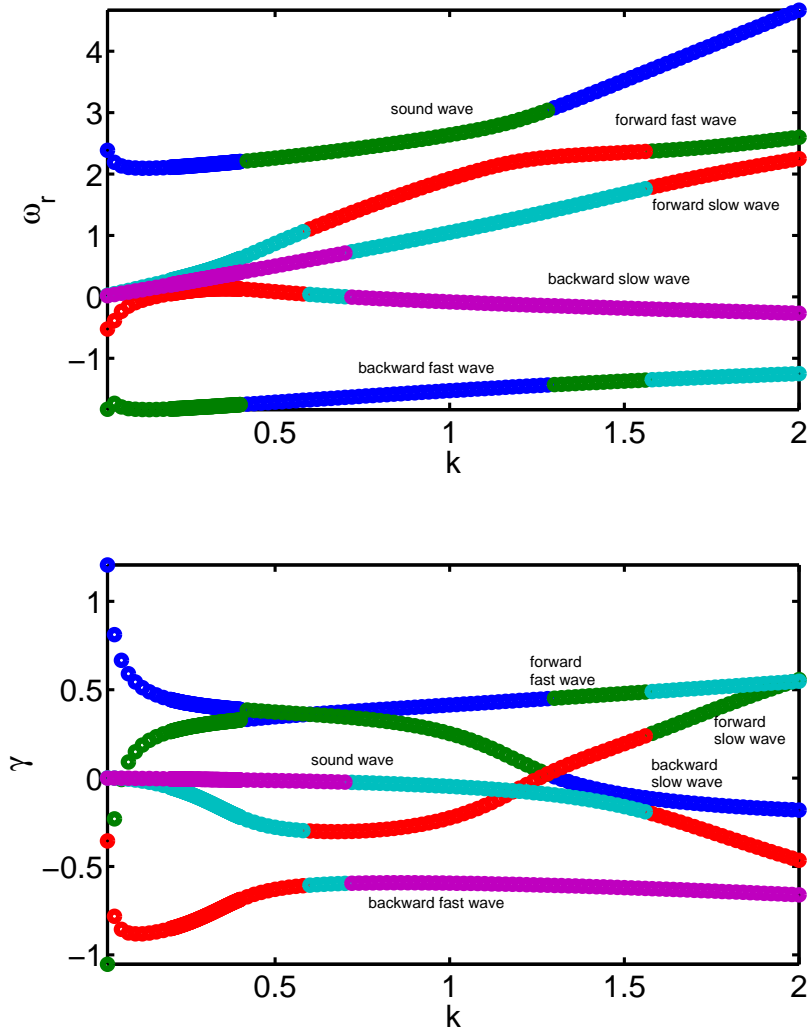}
\caption{Linear instability in the moderately higher $k$ regime. Backward
and forward fast/slow modes are shown respectively by the lower and
upper curves in the upper panel.  Interestingly the modes that decay
in the small $k$ regime, tend to grow eventually when $k$ is order
unity. This is shown in the lower panel. }
\label{fig2}
\end{figure}

We next carry out a linear analysis by assuming fixed (or constant)
background quantities around which the small amplitude perturbed
quantities are expanded. The amplitude of the linear perturbation is
assumed to be smaller compared to the background quantity for a valid
description of the linear theory. All the perturbed variables are
expressed in the form of waves with frequency $\omega$ and wavenumber
$k$. The plasma quantities are described as
\[ \rho(x,t) = \rho_0 +  \tilde{\rho}_k~ e^{ikx-i\omega t},\]
\[ U(x,t) = U_0 +  \tilde{U}_k~ e^{ikx-i\omega t},\]
\[ P(x,t) = P_0 +  \tilde{P}_k~ e^{ikx-i\omega t},\]
\[ B(x,t) = B_0 +  \tilde{B}_k~ e^{ikx-i\omega t},\]
where $\tilde{\rho}_k(x,t), \tilde{U}_k(x,t)\tilde{P}_k(x,t)$ and
$ \tilde{B}_k(x,t)$ are respectively small amplitude density,
velocity, pressure and magnetic field fluctuations.  Note that the
linear frequency $\omega$ is essentially a complex quantity
i.e. $\omega = \omega_r + i \gamma $ where the linear part corresponds
to the wave. By contrast the imaginary part describes growth (if
positive) or damping (if negative) associated with the linear
perturbations. We are interested in determining $\gamma(k)$ in our work.

The corresponding linear expansion for the neutral fluid is
\[ N(x,t) = N_0 +  \tilde{N}_k~ e^{ikx-i\omega t},\]
\[ V(x,t) = V_0 +  \tilde{V}_k~ e^{ikx-i\omega t},\]
\[ P_n(x,t) = P_{n_0} +  \tilde{P_n}_k~ e^{ikx-i\omega t},\]
where $N_0, V_0, P_{n_0}$ are respectively the background constant
neutral density, velocity and pressure, and
$\tilde{N}_k, \tilde{V}_k, \tilde{P_n}_k$ are the amplitude of
corresponding perturbed variables.

On linearizing \eqsto{plasma1}{neut3}, we obtain
\be
-i \omega \tilde{\rho}_k + i k (\rho_0 \tilde{U}_k + U_0 \tilde{\rho}_k)=0,
\ee
\be
-i \omega \rho_0 \tilde{U}_k + \rho_0 U_0 i k \tilde{U}_k 
+ ik P_k = -\nu ( \rho_0 \tilde{U}_k  + \tilde{\rho}_k U_0 -
 \rho_0 \tilde{V}_k - \tilde{\rho}_k V_0),
\ee
\be
-i \omega \tilde{P}_k + i k U_0  \tilde{P}_k + ik P_0 \gamma
\tilde{U}_k= \nu \tilde{\rho}_k
\left[ \frac{\gamma-1}{2} (U_0 -V_0)^2 +v_{th}^2 - \frac{P_0}{2\rho_0} \right],
\ee
\be
\omega = k U_0,
\ee
\be
-i \omega \tilde{N}_k + i k (N_0 \tilde{V}_k + V_0 \tilde{N}_k)=0,
\ee
\be
-i \omega N_0 \tilde{V}_k + N_0 V_0 i k \tilde{V}_k 
+ ik P_{n_k} = \nu ( \rho_0 \tilde{U}_k  + \tilde{\rho}_k U_0 -
 \rho_0 \tilde{V}_k - \tilde{\rho}_k V_0),
\ee
\be
-i \omega \tilde{P}_{n_k} + i k V_0  \tilde{P}_{n_k} + ik P_{n_0} \gamma
\tilde{V}_k= \nu \tilde{\rho}_k
\left[ \frac{\gamma-1}{2} (U_0 -V_0)^2 +v_{th}^2 - \frac{P_{n_0}}{2N_0} \right].
\ee
It is noted that the Alfv\'en wave frequency, $\omega = k U_0$, is
decoupled trivially from the neutral fluid despite charge exchange
coupling interactions. This indicates that Alfv\'en waves are not
affected by the charge exchange interactions. This result can be
elucidated as follows.  Firstly, the charge exchange process tends to
swap the plasma ions with the neutral particles. During this process,
the wave activity is not tampered by the charge exchange. Secondly,
the magnetic field is unperturbed by the charge exchange process and
it continues to remain frozen in the plasma component of the flow.
Hence the linear frequency of Alfv\'en waves continues to remain
unchanged. However, this scenario can change during the nonlinear
evolution in that Alfv\'en waves are shown to exhibit modulation
[Shaikh \& Zank 2010] mitigated by rapid nonlinear growth and damping
that occur alternatively in the coupled plasma-neutral system.

After eliminating all the perturbed variables, we get the fifth order
linear dispersion relation with complex coefficients,
\be
\label{disp}
A_5~ \omega^5 + A_4~ \omega^4 +A_3~ \omega^3 +
A_2~ \omega^2 + A_1~ \omega + A_0 = 0.
\ee
Note further that the sixth order term in the above polynomial exists
separately in the form of the Alfv\'enic branch (i.e.  $\omega = k
U_0$) which is shown to decouple because  the magnetic field, 
critical in exciting Alfv\'en waves, is unaffected largely by the
charge exchange process. The magnetic field fluctuations are embeded
into the plasma component and are affected only indirectly by the plasma
momentum flows. Nonetheless, the complex coefficients in \eq{disp} are
given below;
\[A_5 = N_0, \]
\[A_4 = i \rho_0 (N_0 \nu + \nu)-2 ~ V_0 N_0 k - \rho_0 U_0  N_0 k
-2~U_0  N_0 k,
\]
\eqa
A_3 = i \rho_0 U_0 (-N_0 k \nu - 2 k \nu) + i \rho_0 V_0(-3N_0 k \nu -k \nu) \nonumber \\
+ i\rho_0^2 U_0 k \nu + \gamma (-N_0 P_0 k^2 - P_{n_0}) + \rho_0 
(2~  U_0 V_0 N_0 k  \nonumber \\
+ 2~U_0^2  N_0 k^2) + V_0^2 N_0 k^2 +
4~U_0 V_0 N_0 k^2 + U_0^2 V_0 N_0 k^2,  \nonumber 
\eeq

\eqa
A_2 = \rho_0  N_0 k^2 \nu \delta_p - \rho U_0 V_0^2 N_0k^3 - 4 U_0 V_0 N_0 k^3 - U_0^3 N_0 k^3 \nonumber \\
+ i \rho_0 U_0 V_0 (3N_0 k^2 \nu + 2 k^2 \nu) + i \rho_0 3 V_0^2 N_0 k^2 \nu + \nonumber \\
i \rho_0^2(U_0 V_0  k^2 \nu-2U_0^2 k^2 \nu) + i \rho_0^\nu U_0^2 k^2 ) \nonumber \\
+ \gamma i (-\rho_0 P_0 k^2 \nu - P_{n_0} \rho_0) + 2  \gamma V_0 N_0 P_0 k^3 \nonumber \\
+ \gamma  U_0 N_0 P_0 k^3 +  P_{n_0} (\rho_0 U_0 k + 2 U_0 k) \gamma 
- 2 U_0 V_0^2 N_0 k^3 - 2 U_0^2 V_0 N_0 k^3, \nonumber 
\eeq

\eqa
A_1 = i \rho_0 (-2 V_0 N_0 k^3 \nu \delta_p - V_0^3 N_0 k^3 \nu - 3 U_0 V_0^2 N_0 k^3 \nu - \nonumber \\
U_0^2 V_0 k^2 \nu) + i \rho_0^2 (U_0^3 k^3 \nu - 2 U_0^2 V_0 k^3 \nu) + \rho_0^2 (\nu^2 \delta_p \nonumber \\ 
- k^2 \nu^2 \delta_p) + \gamma i \rho_0 (V_0 P_0 k^3 \nu + U_0 P_0 k^3 \nu) \nonumber \\ 
+ iP_{n_0} \rho_0 (2 U_0 k - \nu U_0 k + V_0 k ) -iV_0^2 N_0 P_0 k^4 -2i U_0 V_0 N_0 P_0  k^4 \nonumber \\
+ iP_{n_0} (-2 \rho_0 U_0^2 k^2 - U_0^2 k^2) + \rho_0 (2 U_0^3 V_0 N_0  k^4+ 2 U_0^2 V_0^2 N_0) \nonumber \\
+ U_0^2 V_0^2 N_0  k^4 + \gamma P_{n_0} P_0, \nonumber 
\eeq

\eqa
A_0 = -\rho_0 U_0^3 V_0 N_0 k^5 - \gamma^2 P_{n_0} U_0 P_0 k \nonumber \\
+ \gamma i P_{n_0} \rho_0 [-k^2 \nu \delta_p + U_0^2 k^2 (\nu-1)-U_0 V_0 k^2] \nonumber \\
- i \rho_0 U_0 V_0  P_0  \nu k^4 + \gamma U_0 V_0^2 N_0 P_0  k^5 + P_{n_0} \rho_0 U_0^3 k^3 \nonumber \\
+ i \rho_0 V_0^2  k^4 \nu \delta_p + i \rho_0 U_0 V_0^3 N_0 k^4 \nu +i \rho_0 U_0^3  V_0  k^4 \nu \nonumber \\
+ \rho_0^2 (V_0 k^3 \nu^2 \delta_p - U_0 k \delta_n \nu^2), \nonumber 
\eeq

\[ \delta_p = \left[ \frac{\gamma-1}{2} (U_0 -V_0)^2 +v_{th}^2 - \frac{P_0}{2\rho_0} \right], \] 
\[ \delta_n =\left[ \frac{\gamma-1}{2} (U_0 -V_0)^2 +v_{th}^2 - \frac{P_{n_0}}{2N_0} \right]. \]

\begin{figure}[t]
\includegraphics[width=11cm]{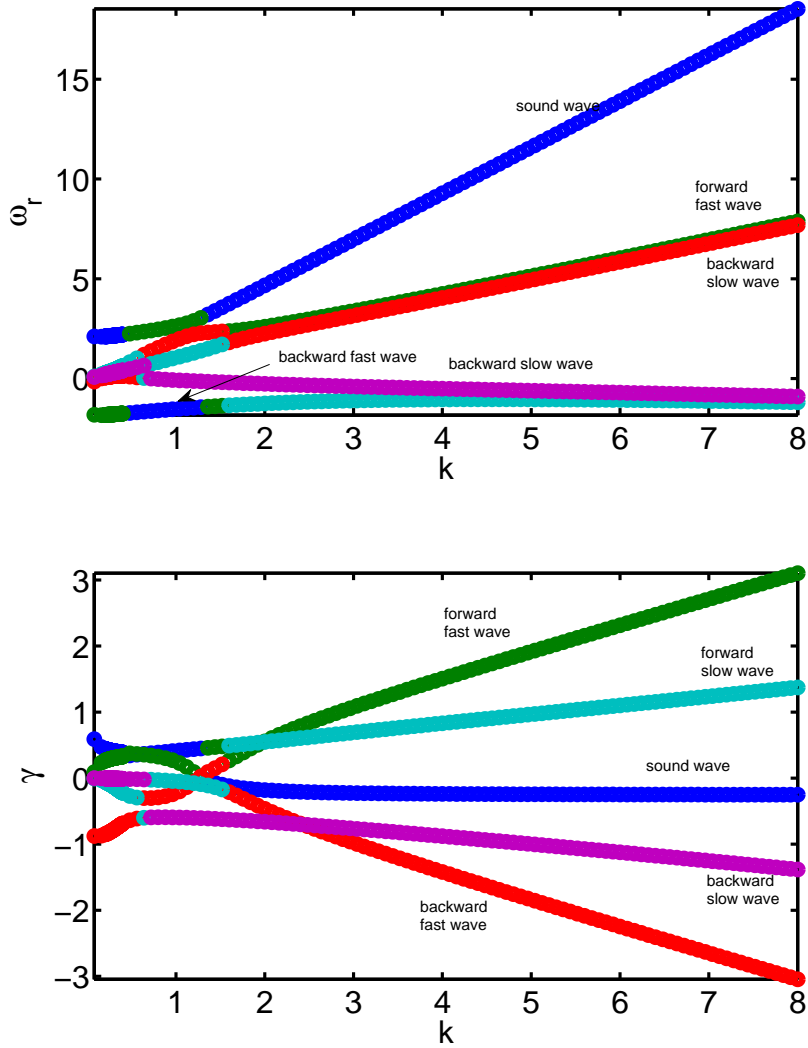}
\caption{Larger modes $k/k_c>1$ (smaller scales) show predominant propagation
of forward fast/slow waves as shown in the upper panel. Smaller scales
show a decoupling with the sound waves. Lower panel shows unstable
modes in the coupled plasma-neutral system for larger modes. Sound
waves are found to be stable for the smaller scales.}
\label{fig3}
\end{figure}

\section{Solution of the dispersion relation}
We next describe the solution of linear dispersion
relation, \eq{disp}, for the coupled plasma-neutral system in the
heliosphere. To access a broader spectrum, we solve \eq{disp} in three
distinct regime, namely the short length scale ($k/k_c>1$), longer
length scale ($k/k_c<1$) and the scales comparable to the charge
exchange length ($k/k_c \simeq 1$), where $k_c = \sqrt{\sigma
n_0}=2\pi/l_c$ is the charge exchange mode ($l_c$ charge scale
length). Charge exchange interactions not only modify the propagating
characteristic of the coupled plasma-neutral modes, but they also
alter the growth and damping rates.

We first examine the affect of charge exchange coupling on the longer
(than charge exchange length $l_c$) length scale part of the spectrum.
The real part of the complex frequency leads to the sound wave,
forward and backward propagating fast/slow magnetosonic waves.  This
is shown in Fig (2).  Alfv\'enic branch is not shown in this figure.
Note that the sound waves in the neutral fluid coexist with the plasma
modes due to the charge exchange coupling. Furthermore, the sound
waves are not destablized by the charge exchange at lower $k$'s.
Interestingly, the charge exchange interactions drive large scales
unstable. It is evident from Fig (2), lower panel, that the linearly
unstable damping modes coexist with the growing (or unstable)
modes. By virtue of this, the mode coupling interactions turn out to
be far more complex. An implication of this instability is to lead to
an alternate growth and damping of the fast/slow waves.  This
instability was thought to govern nonlinear modulation of the
fast/slow wave in the coupled plasma-neutral system reported in Ref
[\cite{dastgeer4}] by us. It is further noted that the sound waves
begin to decouple from the MHD modes for higher $k>1$ modes. This is
an indication that the small scales are not coupled efficiently by the
charge exchange interactions. This is further consistent with our
nonlinear two dimensional simulations [\cite{dastgeer3}] that show a
weak coupling between the small scale turbulent fluctuations and
charge exchange scales.

We next access a moderately higher wavenumber regime in the
spectrum. This regime, depicted in Fig (3), exhibits an entirely
different characteristic compared to the one described in Fig (2).
Clearly, the forward propagating fast and slow magnetosonic modes
dominate over the slow (below zero roots) ones in the higher $k$
spectrum where $k>k_c$ (but not $k \gg k_c$). By contrast, the
unstable modes (lower panel in Fig 3) exhibits interesting feature. In
that, the linearly unstable fast/slow modes (i.e the second curve from
the top) shows initial growth (i.e $\gamma >0$) for $k<1.25$. This
characteristic behavior reverses for $k>1.25$. Converse is true for
the third (from the top) curve. On the other hand, the top and lowest
curves corresponding to the forward and backward propagating fast/slow
modes show respectively the growth and damping of the instability.

Finally, to investigate the asymptotic behavior of these linearly
stable/unstable modes, we extend our analysis to span the higher $k$
modes in the spectrum. This is shown in Fig (4). It becomes clear from
this figure that the linear wave propagation (upper panel) show a
considerable shift in the frequency towards the forward propagating
branch ($\omega_r>0$), whereas the frequency associated with the
backward propagation is found to be reduced considerably.
Interestingly, the fast/slow branches associated with the magnetosonic
waves appear to have merged for the smaller scales in the
$k$-spectrum.  The unstable modes, (Fig 4, lower panel), on the other
hand exhibit a predictable behavior. It is found that the forward
propagating fast/slow modes are driven unstable by the charge exchange
coupling in the coupled plasma-neutral system. On the other hand, the
backward propagating fast/slow modes are damped by the charge exchange
interactions. This is described by the lower panel in Fig (4). It is
further observed that the sound waves, in the higher $k$ regime, are
decoupled from the predominant MHD modes. This means that charge
exchange do not mix up these modes at relatively small characteristic
length scales.

\section{Summary and conclusion}
A major outcome from our analysis is that charge exchange interactions
in the partially ionized gases, dominated by the coupled
plasma-neutral system, in the heliosphere {\it do not directly modify}
the propagation characteristic of the linear Alfv\'en waves.  This
does not necessarily mean that their nonlinear evolution is also
unaffected.  As a matter of fact, the nonlinear evolution of Alfv\'en
waves is influenced by their mode coupling interaction with the
fast/slow magnetosonic waves. This is discussed by Shaikh \& Zank
(2010). We however find that the charge exchange interaction influence
fast/slow mode by linearly driving them unstable.

The linear analysis described in the paper is extremely useful in
explaining some of the features of our nonlinear simulations
[\cite{dastgeer3,dastgeer4}]. The latter show that charge exchange
modes modify the helioshperic turbulence cascades dramatically by
enhancing nonlinear interaction time-scales on large scales.  By
contrast, small scale turbulent fluctuations are unaffected by charge
exchange modes which evolve like the uncoupled system as the latter
becomes less important near the larger $k$ part of the helioshperic
turbulent spectrum.  This tends to modify the characteristics of
helioshperic plasma turbulence which can be significantly different
from the Kolmogorov phenomenology of fully developed turbulence.

The onset of modulational instability in Ref [\cite{dastgeer4}] is
triggered essentially by linear instability process.  Our linear
stability analysis indicates that the underlying coupled
plasma–neutral system possesses fast/slow unstable modes. These modes
account for the growing as well as damping of Alfvénic and fast/slow
compressive waves. The latter leads to a concurrent damping of the
fast/slow compressive mode preceding the growth of Alfv\'enic mode
occurs. When the two modes reach their extremal rise or fall, they
reverse their behavior. The fast/slow compressive mode begins to rise
at the expense of damping of Alfv\'enic mode. This process continues
nonperiodically and repeats itself in time thereby exhibiting a nearly
oscillatory behavior. It then follows that the two modes,
i.e. Alfv\'enic and fast/slow compressive, regulate each other in a
predator–prey manner. Dynamically, the Alfv\'enic mode grows at the
expense of fast/slow mode. When Alfv\'enic mode reaches its maximum
amplitude, where the fast/slow mode remains at its lowest magnitude,
the latter eats up the Alfv\'enic mode and vice versa. This process is
repetitive.

Our results should find application to a variety of astrophysical
environments in which a partially ionized plasma is typical. Examples
include the outer heliosheath formed by the interactions of the solar
wind with the local interstellar medium, the magnetic collapse of
molecular clouds and star formation and the general transfer of energy
in partially ionized plasmas surrounding other stellar systems
[\cite{wood}].

\section{Acknowledgment}
This work was performed under PH499 capstone course during the summer
2010.  The partial support of NASA grants NNX09AB40G, NNX07AH18G,
NNG05EC85C, NNX09AG63G, NNX08AJ21G, NNX09AB24G, NNX09AG29G, and
NNX09AG62G is acknowledged.



\begin{thereferences}{19}

\bibitem{balsara}
Balsara, D., ApJ., 1996, 
465, 775.


\bibitem{fite}
Fite, W. L., Smith, A. C. H., and Stebbings, R. F., 1962,
Proc. R. Soc. London, em A. 268, 527.

\bibitem{Florinski2003}
Florinski, V., G. P. Zank, and N. V. Pogorelov (2003),
J. Geophys. Res., 108(A6), 1228, doi:10.1029/2002JA009695.

\bibitem{Florinski2005}
Florinski V., G. P. Zank, N. V. Pogorelov (2005), J. Geophys. Res.,
110, A07104, doi:10.1029/2004JA010879.

\bibitem{lab} 
Gigliotti, A., Gekelman, W., Pribyl, P.,  Vincena, S.,
Karavaev, A., Shao, X., Sharma, A.,  Papadopoulos, D., 2009,
  Phys. Plasmas,  16, 092106.


\bibitem{Gekelman}
Gekelman, W., 2004,
Journal of Geophysical Research,  109,  A01311.


\bibitem{kulsrud}
Kulsrud, R., and Pearce, 1969, 
ApJ 156, 445.



\bibitem{leake}
Leake, J.,  Arber,  T. D., M. L. Khodachenko,  M. L., 2005,
Astronomy and Astrophysics, 442, 1091.


\bibitem{oishi}
 Oishi, J. S. and  Mac Low, M., 2006,
 ApJ, 638, 281.



\bibitem{padoan}
Padoan, P.,  Zweibel, E., and  Nordlund, A., 2000,
ApJ, 540, 332.

\bibitem{pauls1995}
Pauls, H. L., Zank, G. P., and Williams, L. L., 1995, 
J. Geophys. Res. A11, 21595.


\bibitem{dastgeer} 
Shaikh, D., and  Zank, G. P., 2008,
The Astrophysical Journal, Volume 688, Issue 1, pp. 683-694.

\bibitem{dastgeer2} 
Shaikh, D., and  Zank, G. P., 2006,
The Astrophysical Journal, Volume 640, Issue 2, pp. L195-L198.

\bibitem{dastgeer3} 
Shaikh, D. and Shukla, P. K., 2009, Physical Review Letters, vol. 102,
Issue 4, id. 045004

\bibitem{dastgeer4} 
Shaikh, D., and  Zank, G. P., 2006,
Physics Letters A 374 (2010) 4538-4542.

\bibitem{wood} Wood, B. E., Harper, G. M.;
Muller, H. R., Heerikhuisen, J., Zank, G. P., ApJ, 655, 946, 2007.

\bibitem{zank1999}
 Zank,  G. P., 1999, 
Space Sci. Rev., 89, 413-688, 1999.


\end{thereferences}
\end{document}